\documentclass[10pt]{iopart}
\usepackage{iopams}
\usepackage{graphicx}
\usepackage{vrsion}

\begin{document}

\newcommand{\degc}{\ensuremath{^{\circ}}C}

\bibliographystyle{phcpc}

\title{Equilibrium cluster formation and gelation}

\author{ Rodrigo Sanchez and Paul Bartlett}

\address{School of Chemistry, University of Bristol, Bristol BS8 1TS, UK}

\ead{P.Bartlett@bristol.ac.uk}

\begin{abstract}
We study the formation and growth of equilibrium clusters in a
suspension of weakly-charged colloidal particles and small
non-adsorbing polymers. The effective potential is characterised by
a short-range attraction and a long-range repulsion. The size, shape
and local structure of the clusters are studied using
three-dimensional particle microscopy. We observe
a rapid growth in the mean cluster size and the
average number of nearest neighbours approaching the gel boundary.

\end{abstract}


\submitto{\JPCM}


\section{Introduction}

The van der Waals theory is the foundation for our current
understanding of fluid structure and phase equilibria
\cite{Widom-771}. It predicts, for an interaction potential
consisting of a short-range repulsion and a long-range attraction, a
first-order transition (below $T_{c}$) between a dilute vapour and a
dense fluid or liquid. Here we consider the unusual situation which
pertains when the range of the repulsive and attractive forces are
 interchanged.

Experiments \cite{2233,3338,3357,3712,3514}, theory
\cite{3267,3298,3502,3501} and simulation \cite{3276,3247,3672}
suggest that a combination of a very short-ranged attraction and a
much longer range repulsion leads to a highly nontrivial phase
behaviour. At low densities, there is compelling evidence for the
existence of a fluid phase of finite-sized \textit{equilibrium}
aggregates or clusters \cite{2233,3338,3357}. The size and shape of
these clusters are expected \cite{3247} to be very sensitive to the
relative range of the attractive and repulsive components
($V_{R}(r)$) of the total interparticle potential. Estimates of the
ground state energies of clusters of different sizes suggest that
the equilibrium shape changes from spherical to linear as the
cluster diameter exceeds the  decay length of $V_{R}(r)$
\cite{3247,3672}. In addition recent experiments \cite{3514} have
revealed that, at high densities, the competition inherent in this
system can also stabilize new particle architectures.
Three-dimensional confocal microscopy in a system of weakly charged
colloidal spheres show, at high densities, the formation of a
macroscopically percolating gel composed of linear aggregates of
face-sharing tetrahedra of particles. The colloidal particles are
arranged in the motif of a helical spiral in which each particle is
connected to six neighbours, analogous to the structure first
proposed by Bernal in his classic study of hard-sphere glasses
\cite{407}.

In this paper, we present an experimental study of a
carefully-characterized model system of colloidal spheres. The
particles interact through an effective potential which consists of
a short-range attraction, generated by the exclusion of a
non-adsorbing polymer, and a long-range screened electrostatic
repulsion.  We focus on what happens to the size and shape of the
self-assembled clusters as the gel boundary is approached.

\section{Experimental details}

 We use a well-studied model system:
random-coil polystyrene (radius of gyration $r_{g} = 92 $ nm) mixed
with PMMA spheres (diameter $\sigma=1700$ nm) dispersed in a
near-density and near-index matched mixture of 78:22 \% by weight
cycloheptyl bromide and \textit{cis}-decalin. The PMMA spheres were
labelled with the dye DiIC$_{18}$ to make them visible by
fluorescence microscopy. Exclusion of the non-adsorbing polymer from
a region between two neighbouring particles generates a short-range
attraction. The range and strength of which is controlled by $r_{g}$
and the free polymer concentration $\phi_{p}^{*}$. For the cases
studied here the width of the attractive range is roughly $0.1
\sigma$. Measurements of the particle electrophoretic mobility using
phase analysis light scattering revealed that the PMMA particles
develop a small positive charge, $Q \approx +145 e$, in the
density-matched solvent used in this study.  The Debye screening
length was estimated from measurements of the solvent conductivity
($170 \pm 30$ pS cm$^{-1}$) as of the order of $\kappa \sigma
\approx 0.5$.

\begin{figure}
\begin{center}
  \includegraphics[width=3.0in]{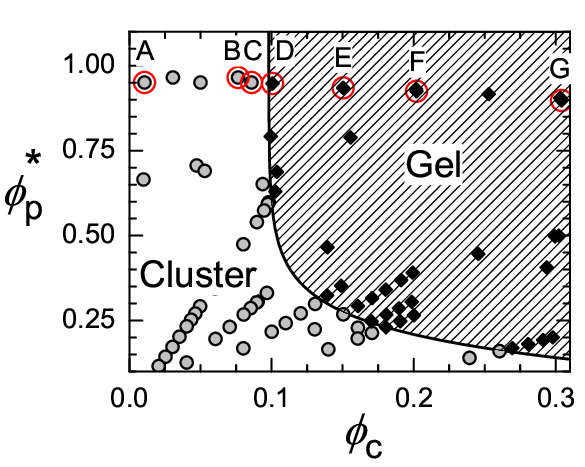}\\
  \caption{Behaviour of a charged colloid-polymer mixture. Circles represent
  samples that form isolated clusters. Diamonds depict macroscopic percolating gels. The solid line identifies
  the experimentally-determined gel boundary $\phi_{g}$.
  Data from circled samples A--G
  are shown in \fref{fig:cluster-prop}--\ref{fig:cluster-asp}.} \label{fig:phase}
\end{center}
\end{figure}

Mixtures with different colloid ($\phi_{c}$) and polymer
concentrations ($\phi_{p}^{*}$) were prepared and homogenized by
extensive tumbling. To minimize decomposition and to ensure
reproducibility all samples, once prepared, were stored in
light-tight containers at 4\degc, prior to measurement. The size and
internal structure of clusters were studied using laser scanning
confocal microscopy.  The suspensions were sealed in a cylindrical
cell of $\sim 50$ $\mu$L volume. Three-dimensional confocal
microscopy was used to distinguish between a phase of isolated
clusters and the formation of a macroscopic percolating network of
particles. Our observations are summarized in \fref{fig:phase}. To
follow the transition between isolated clusters and a gel phase in
detail  we focus here on the sample sequence A--G, \fref{fig:phase},
at  $\phi_{p}^{*} = 0.95$. A stack of typically 345 images, spaced
by 0.16 $\mu$m vertically, was collected from a representative
volume of 73 $\mu$m x 73 $\mu$m x 55 $\mu$m within each suspension.
Each stack contained, depending on $\phi_{c}$, the images of between
1500 and 45000 spheres. The images were analyzed quantitatively
using algorithms similar to those described in \cite{Crocker-966} to
extract the three-dimensional coordinates of each sphere. The
position of each sphere was measured with a precision of
approximately 70 nm.

\section{Data analysis}

The number, size and shape of the clusters were
identified from the coordinate list. We define particles which are
`bonded' to each other by their separation $r$. If $r \leq r_{0}$,
where the cut-off distance $r_{0}$ was identified with the position
of the first minimum in the pair distribution function $g(r)$, then
the particles were considered neighbours. Standard algorithms were
used to partition particles into clusters of size $s$ and to
evaluate the cluster size distribution $n_{s}$. The second moment of
the cluster distribution $\langle s_{2}\rangle$ was used to
characterize the average cluster size \cite{3713},
\begin{equation}\label{eq:s2}
\langle s_{2} \rangle = \frac{\sum_{s} s^{2}n_{s}}{\sum_{s} sn_{s}}.
\end{equation}
The geometric properties of the clusters were evaluated from the
eigenvalues $\lambda_{1} \leq \lambda_{2} \leq \lambda_{3}$ of the
radius of gyration tensor $\bi{S}$, defined as
\begin{equation}\label{eq:rgtensor}
    S_{\alpha \beta} = \frac{1}{2N^{2}} \sum_{i,j = 1}^{N}
    [r_{i,\alpha} - r_{j, \alpha}][r_{i,\beta} - r_{j, \beta}]
\end{equation}
with $\bi{r}_{i}$ the position of the $i$th sphere in the cluster of
size $N$ and $(\alpha, \beta) = $ 1, 2, 3 denoting the corresponding
Cartesian components. We characterize the size and shape of the
cluster in terms of the rotational invariants of the tensor
$\bi{S}$. The trace $\sum_{i} \lambda_{i}$ gives the usual isotropic
squared radius of gyration $r_{g}^{2}$ while the anisotropy of the
cluster is measured by the asphericity $A_{2}$ \cite{3692},
\begin{equation}\label{eq:asphericity}
A_{2}  = \frac{(\lambda_{1}-\lambda_{2})^{2} +
(\lambda_{2}-\lambda_{3})^{2} +
(\lambda_{3}-\lambda_{1})^{2}}{2(\lambda_{1}+\lambda_{2}+\lambda_{3})^{2}}.
\end{equation}
The asphericity is normalized so that $A_{2}$ ranges from zero for a
spherically symmetric cluster to one for a rod-like cluster.

\section{Results}

\begin{figure}
\begin{center}
  \includegraphics[width=3.5in]{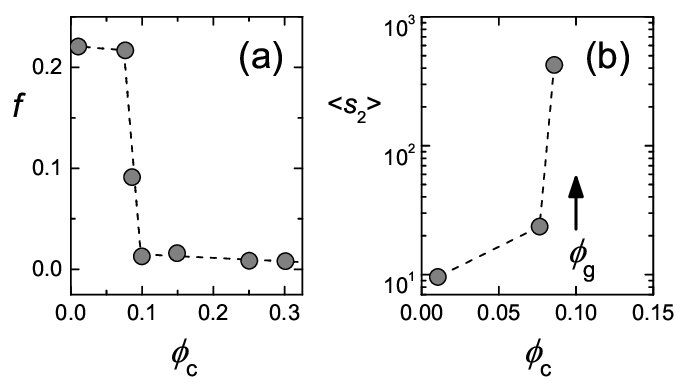}\\
  \caption{(a) The proportion of unbonded particles $f$ and (b) the average cluster
  size $\langle s_{2} \rangle$,  as a function of colloid volume fraction $\phi_{c}$, for samples marked
   A--G in \fref{fig:phase}.} \label{fig:cluster-prop}
\end{center}
\end{figure}

The left panel of \fref{fig:cluster-prop} shows the evolution in the
proportion of \textit{single} unbonded particles $f$ as a function
of $\phi_{c}$. The boundary between a non-percolating cluster phase
and the percolating gel occurs at $\phi_{g} \approx 0.1$. The most
striking feature is the abrupt change in the number of single
particles approaching the gel boundary. At low colloid
concentrations $f$ is approximately constant at $f \approx 0.22$,
independent of $\phi_{c}$ - the remaining 78\% of particles being
incorporated into clusters. The proportion of single particles
however drops rapidly in the gel phase, although it remains finite
even for dense gels ($f \approx 10^{-2}$ at $\phi_{c} = 0.3$, for
instance). The coexistence evident in these samples between free and
bonded particles highlights the dynamic nature of the assembly
process. Visual observation revealed particles continuously joining
and leaving clusters. To provide further evidence that equilibrium
had been reached we studied the time-dependence of $f$. We found no
qualitative change on the period of a week, during which time
particle sedimentation was negligible.

\begin{figure}
\begin{center}
  \includegraphics[width=4.5in]{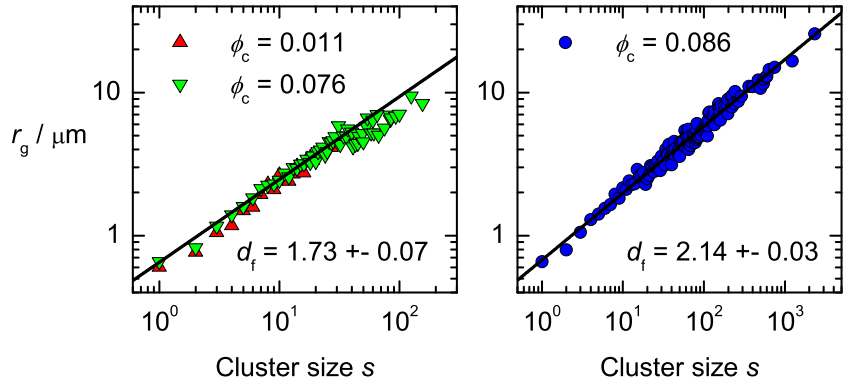}\\
  \caption{Size dependence of the cluster radius of gyration:
  sample A, $\phi_{c} = 0.011$ (triangles), sample B, $\phi_{c} = 0.076$ (inverted triangles),
  and sample C, $\phi_{c} = 0.086$ (circles). The lines are power law fits to the data.} \label{fig:rg}
\end{center}
\end{figure}

\Fref{fig:cluster-prop}(b) shows the growth in the average cluster
size $\langle s_{2} \rangle$, defined in \eref{eq:s2}, with
increasing $\phi_{c}$ for all non-percolating states. Clearly the
the average cluster size increases rapidly approaching the gel
boundary. This rapid growth is mirrored in the number of particles
$s_{max}$ contained in the largest observed cluster, which increases
from 37 ($\phi_{c} = 0.011$), to 155 ($\phi_{c} = 0.076$), to reach
$s_{max}=2340$ at  $\phi_{c} = 0.086$. To quantify these changes in
greater detail we determined the size dependence of the
ensemble-averaged radius of gyration $\langle r_{g} \rangle $. For
fractal aggregates $r_{g}$ displays a power law dependence on the
number of particles $s$, $\langle r_{g} \rangle \sim s^{1/d_{f}}$,
where $d_{f}$ is the fractal dimension. \Fref{fig:rg} shows plots of
$r_{g}$ versus $s$, at three values of $\phi_{c}$. Interestingly we
find that at all densities the clusters show fractal scaling but
that the fractal dimension $d_{f}$ shows a marked dependence on
$\phi_{c}$. At low $\phi_{c}$, the fractal dimension found of $d_{f}
= 1.73 \pm 0.07$ is consistent with the diffusion-limited
cluster-cluster aggregation (DLCA) value in three dimensions ($d_{f}
= 1.78$ \cite{2937,3714}). While at high $\phi_{c}$, the value of
$d_{f} = 2.14 \pm 0.03$ is consistent with the reaction-limited
cluster-cluster aggregation (RLCA) value in three dimensions ($d_{f}
= 2.1$ \cite{Lin-785}). This apparent crossover between DLCA and
RLCA kinetics with $\phi_{c}$ is unexpected.  However it is unclear
if fractal models for non-equilibrium growth are applicable to the
equilibrium case of assembly studied here for which, as we indicate
below, the cluster shape is not truly self-similar.

\begin{figure}
\begin{center}
  \includegraphics[width=5.0in]{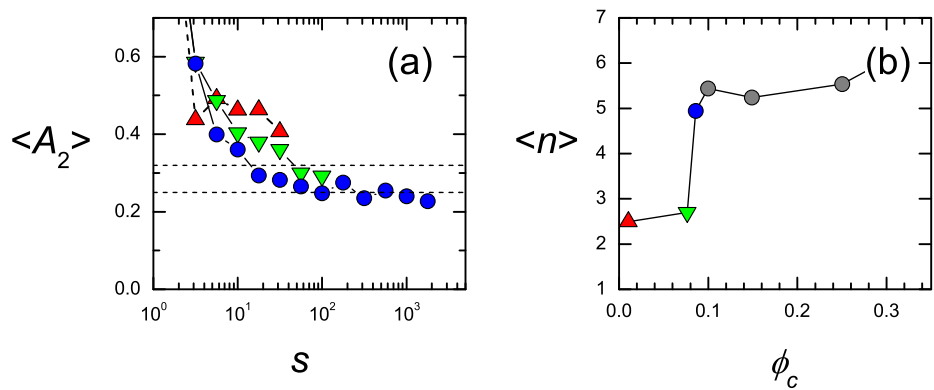}\\
  \caption{(a) Cluster asphericity as a function of cluster size. The dotted lines depict
  the asphericity expected for random percolation \cite{3696} and DLCA models \cite{3695}.
   (b) Average number of neighbours as a function
  of volume fraction $\phi_{c}$. Symbols have the same meaning as \fref{fig:rg}.} \label{fig:cluster-asp}
\end{center}
\end{figure}

A simple indicator of the anisotropy of the clusters is provided by
the  asphericity $\langle A_{2} \rangle$, defined in
\eref{eq:asphericity}. In contrast to the $r_{g}$ data we find
little $\phi_{c}$ dependence of the average cluster asphericity,
plotted in \fref{fig:cluster-asp}(a). Both low and high density
samples show a similar monotonic decrease in $\langle A_{2} \rangle$
with $s$, consistent with a progressive change in shape with
increasing cluster size. The clusters formed are therefore not
self-similar. Clusters smaller than about 30--50 monomers show a
definite tendency to be elongated in one dimension as evidenced by
the increasing value of $\langle A_{2} \rangle$ found at small $s$.
This preference for a linear or rod-like shape seems to be more
pronounced the smaller the cluster. These observations are supported
by images of individual clusters which reveal characteristic
one-dimensional bundles of particles. By contrast, while large
clusters remain anisotropic the levels of asymmetries are
significantly lower and typical of clusters formed from either
random percolation ($\langle A_{2} \rangle = 0.25$ \cite{3696}) or
DLCA models $\langle A_{2} \rangle = 0.32$ \cite{3695}). This
crossover suggests that in large clusters the short one-dimensional
bundles have branched many times generating clusters with geometries
controlled by random percolation.

The tendency seen for one-dimensional aggregation at short scales
might indicate that the cluster is constructed locally from segments
of Bernal spiral structures. To check for this possibility, we
calculated the average number of nearest neighbours $\langle n
\rangle$. As shown in the right hand panel of
\fref{fig:cluster-asp}, $\langle n \rangle$ grows abruptly with
increasing $\phi_{c}$ approaching the gel boundary. Above
$\phi_{g}$, the mean coordination number is insensitive to
$\phi_{c}$ and is approximately six, consistent with the
characteristic six-fold coordination of a Bernal spiral. Moreover
\fref{fig:cluster-asp} reveals that $\langle n \rangle $  is less
than six in the non-percolating cluster samples (A--C). This is
particularly evident in sample B ($\phi_{c} = 0.076$) which while it
contains clusters of upto 100 monomers has a mean coordination
number of only $\langle n \rangle = 2.7$. The lack of spiral
ordering, shown by the low values of $\langle n \rangle$, is also
supported by direct inspection which reveals little evidence for
Bernal ordering in any of the non-percolating cluster samples.

\section{Conclusions}

In this work we have explored the change in the size and shape of
the clusters formed in a colloidal system of short range attractions
and long range repulsions, approaching the gelation transition. In
contrast to a purely attractive system the clusters are equilibrium
entities and their properties time-independent. Our most noteworthy
finding is that the average size  $\langle s_{2} \rangle$ of the
cluster distribution, the proportion of unattached particles, and
the mean coordination number all change abruptly as $\phi
\rightarrow \phi_{g}$. Although we have studied only a limited
number of concentrations (at one specific choice of potential
parameters) these observation suggest that, for this system,
gelation has many of the hallmarks of an \textit{equilibrium}
transition. Qualitatively our observations appear consistent with
the suggestion by Groenewold and Kegel \cite{3267,3298} that under
the right conditions a fluid of isolated clusters will undergo a
spinodal decomposition, driven by counter-ion condensation, to form
a macroscopic gel. This raises the enticing possibility of an
"ideal" thermally-reversible gel -- one in which the gel-line might
be approached infinitely closely and the dynamics of arrest studied
precisely and reproducibly, free from the dynamical complications of
purely attractive systems. Further experiments are planned to test
these ideas.

\ack

It is a pleasure to thank Drs A I Cambell, J S van Duijneveldt and M
Faers for assistance. This work was supported financially by EPSRC
and MCRTN-CT-2003-504712.


\section*{References}
\begin{thebibliography}{10}

\bibitem{Widom-771}
Widom B
\newblock 1967 \textit{Science} {\bf 157} 375

\bibitem{2233}
Segre P~N, Prasad V, Schofield A~B and Weitz D~A
\newblock 2001 \textit{Phys. Rev. Lett.} {\bf 86}  6042

\bibitem{3338}
Sedgwick H, Egelhaaf S and Poon W C~K
\newblock 2004 \textit{J. Phys.: Condens. Matter} {\bf 16}  S4913

\bibitem{3357}
Stradner A, Sedgwick H, Cardinaux F, Poon W C K, Egelhaaf S and
Schurtenberger P
\newblock 2004 \textit{Nature} {\bf 432}  492

\bibitem{3712}
Bordi F, Cametti C, Diociaiuti M and Sennato S
\newblock 2005 \textit{Phys. Rev. E} {\bf 71} 050401(R)

\bibitem{3514}
Campbell A~I, Anderson V~J, van Duijneveldt J~S and Bartlett P
\newblock 2005 \textit{Phys. Rev. Lett.} {\bf 94} 208301

\bibitem{3267}
Groenewold J and Kegel W~K
\newblock 2001 \textit{J. Phys. Chem. B} {\bf 105} 11702

\bibitem{3298}
Groenewold J and Kegel W~K
\newblock 2004 \textit{J. Phys.: Condens. Matter} {\bf 16}  S4877

\bibitem{3502}
Wu J, Liu Y, Chen W-R, Cao J and Chen S-H
\newblock 2004 \textit{Phys. Rev. E} {\bf 70} 050401(R)

\bibitem{3501}
Liu Y, Chen W-R and Chen S-H
\newblock 2005 \textit{J. Chem. Phys.} {\bf 122} 044507

\bibitem{3276}
Sciortino F, Mossa S, Zaccarelli E and Tartaglia P
\newblock 2004 \textit{Phys. Rev. Lett.} {\bf 93} 055701

\bibitem{3247}
Mossa S, Sciortino F, Tartaglia P and Zaccarelli E
\newblock 2004 \textit{Langmuir} {\bf 20} 10756

\bibitem{3672}
Sciortino F, Tartaglia P and Zaccarelli E
\newblock 2005 \textit{preprint} cond-mat/0505453

\bibitem{407}
Bernal J~D
\newblock 1964 \textit{Proc. Roy. Soc. (London) A} {\bf 280} 299

\bibitem{Crocker-966}
Crocker J~C and Grier D~G
\newblock 1996 \textit{J. Coll. Interf. Sci.} {\bf 179} 298

\bibitem{3713}
Stauffer D
\newblock 1985 \textit{Introduction to Percolation Theory}
\newblock (London: Taylor and Francis)

\bibitem{3692}
Rudnick J and Gaspari G
\newblock 1986 \textit{J. Phys. A: Math. Gen.} {\bf 19} L191

\bibitem{2937}
Weitz D~A, Huang J~S, Lin M~Y and Sung J
\newblock 1985 \textit{Phys. Rev. Lett.} {\bf 54} 1416

\bibitem{3714}
Meakin P
\newblock 1989 \textit{J. Coll. Interf. Sci.} {\bf 102} 491

\bibitem{Lin-785}
Lin M~Y, Lindsay H M,  Weitz D A,  Ball R C, Klein R and Meakin P
\newblock 1990 \textit{Phys. Rev. A} {\bf 41} 2005

\bibitem{3696}
Aronovitz J~A and Stephens M~J
\newblock 1987 \textit{ J. Phys. A: Math. Gen.} {\bf 20} 2539

\bibitem{3695}
Fry D, Mohammad A, Chakrabarti A and Sorensen C~M
\newblock 2004 \textit{Langmuir} {\bf 20} 7871

\end{thebibliography}

\end{document}